\documentclass[fleqn]{annalen}
\input{epsf}
\begin{document}
\newcommand{\volume}{}              
\newcommand{\xyear}{1999}            
\newcommand{\issue}{}               
\newcommand{\recdate}{21 July 1999}  
\newcommand{\revdate}{dd.mm.yyyy}    
\newcommand{\revnum}{0}              
\newcommand{\accdate}{dd.mm.yyyy}    
\newcommand{\coeditor}{ue}           
\newcommand{\firstpage}{1}         
\newcommand{\lastpage}{6}          
\setcounter{page}{\firstpage}        
\newcommand{\keywords}{2$d$ metal-insulator transition, 
Anderson localization, interacting fermions} 
\newcommand{\PACS}{71.30.+h, 72.15.Rn, 71.27.+a}
\newcommand{\shorttitle}{X. Waintal et al., Metallic  
phase in $2d$} 
\title{Metallic phase between the Fermi glass \\ and the Wigner crystal in
two dimensions}  
\author{X.\ Waintal$^{1}$, G.\ Benenti$^{1}$, and J.-L.\ Pichard$^{1}$} 
\newcommand{\address}
  {$^{1}$CEA, Service de Physique de l'Etat Condens\'e,\\   
  Centre d'Etudes de Saclay, F-91191 Gif-sur-Yvette, France} 
\newcommand{\email}{\tt pichard@spec.saclay.cea.fr} 
\maketitle
\begin{abstract}
For intermediate Coulomb energy to Fermi energy ratios $r_s$,  
spinless fermions in a two-dimensional random potential form a 
new quantum phase, different from the Fermi glass (weakly  
interacting Anderson localized states) and the Wigner crystal 
(regular array of charges pinned by the disorder). 
The intermediate phase is characterized by an ordered flow of 
persistent currents with a typical value decreasing 
with excitation energy. 
Extending finite size scaling analysis to the many body ground state, 
we find that electron-electron interactions can drive the Fermi glass 
towards an intermediate metallic phase (Coulomb metal). 
\end{abstract} 

\section{Introduction}

According to the scaling theory of localization \cite{abrahams},
all states are localized for non-interacting particles in 
two dimensions ($2d$). 
This absence of a metallic state in $2d$ is nowadays challenged 
since a lot of transport measurements, following the pioneering 
experiments of Kravchenko and co-workers \cite{kravchenko}, 
give evidence of an insulator-metal transition (IMT) when  
the carrier density is increased. 
Many experimental results suggest that Coulomb repulsion drives  
this phenomenon: 
(1) In a very clean heterostructure the IMT was observed 
at a Coulomb energy to Fermi energy ratio  
$r_s\approx 35$ \cite{yoon}, close to $r_s\approx 37$ for which    
a clean Wigner crystal melts according to Monte Carlo simulations
\cite{tanatar};   
(2) In more disordered samples, the IMT typically occurs 
at $r_s\approx 10$, again compatible with the value expected for 
the melting of a pinned Wigner crystal \cite{chui}; 
(3) At a weaker $r_s$ interactions no longer dominate and a 
re-entry towards an insulating phase (Fermi glass of weakly 
interacting localized particles) was observed at $r_s\approx 6$ in 
\cite{hamilton}. 
As at the IMT $k_F l \approx 1$ ($k_F$ denoting the Fermi wave vector
and $l$ the elastic mean free path), disorder is important and has to  
be considered, together with interactions, in a non perturbative way.   

Taking advantage of numerical techniques, 
we consider a model of $N$ Coulomb interacting spinless fermions 
in a disordered square lattice with $L^2$ sites. 
The Hamiltonian reads: 
\begin{equation} 
\label{hamiltonian} 
H=-t\sum_{<i,j>} c^{\dagger}_i c_j +  
\sum_i v_i n_i +U \sum_{i\neq j} \frac{n_i n_j}{2 r_{ij}},  
\end{equation} 
where $c^{\dagger}_i$ ($c_i$) creates (destroys) an electron in 
the site $i$, $t$ is the strength of the hopping terms 
between nearest neighbours ($t=1$ in the following) and $r_{ij}$ 
the inter-particle distance for a $2d$ torus.
The random potential 
$v_i$ of the site $i=(i_x,i_y)$ with occupation number 
$n_i=c^{\dagger}_i c_i$ 
is taken from a box distribution of width $W$. The interaction strength 
$U$ yields $r_s=U/(2t\sqrt{\pi n_e})$ for a filling factor $n_e=N/L^2$. 
A Fermi golden rule approximation for the elastic scattering time   
leads, for $n_e\ll 1$, to $k_F l \approx 192 \pi n_e (t/W)^2$.    

\section{Persistent currents}  

\begin{figure}
\centerline{\epsfxsize=13cm\epsfysize=13cm\epsffile{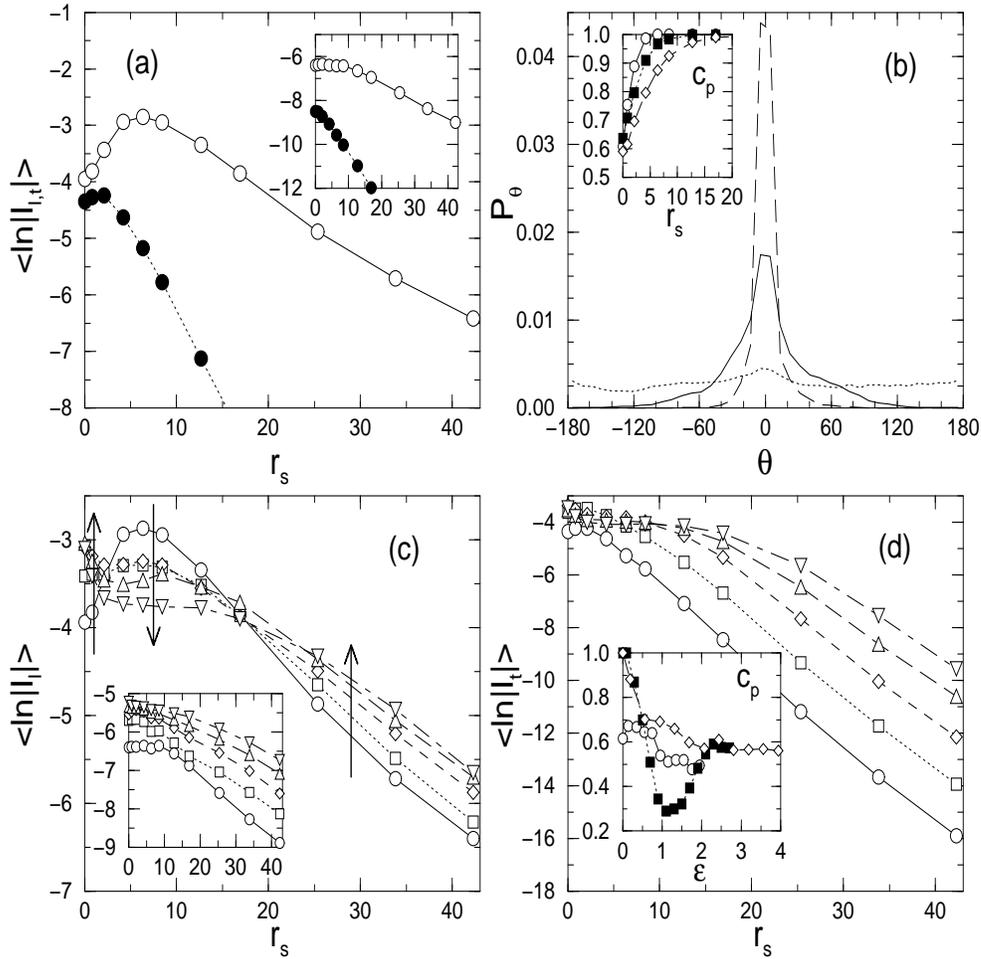}} 
\caption{
(a) Log-averages of the ground state longitudinal current $I_l$ 
(open circles) and transverse current $I_t$ (filled circles) 
at $W=5$. 
Insert: the same at $W=15$. 
(b) Distribution $P_\theta$ of the ground state local currents  
angles at $W=5$, for $r_s=0$ (dotted line), $r_s=6.3$ (full line), 
and $r_s=42$ (dashed line). 
Insert: fraction $c_p$ of paramagnetic samples for $W=5$
(circles), $W=10$ (squares) and $W=15$ (diamonds). 
(c) Log-average of the longitudinal current at $W=5$, for the 
ground state (circles), the first excited state (squares), the 
second-fourth excited states (diamonds), the fifth-ninth excited 
states (triangles up), and the tenth-nineteenth excited states 
(triangles down). Arrows indicate increase (or decrease) of the 
longitudinal current with excitation energy. 
Insert: the same at $W=15$. 
(d) Same as in (c) but for the transverse current at $W=5$. 
Insert: fraction of paramagnetic samples as a function of the 
excitation energy $\epsilon$ at $W=5$, for $r_s=0$ (circles),
$r_s=6.3$ (squares), and $r_s=42$ (diamonds). 
}   
\label{fig1}
\end{figure}

To measure delocalization effects induced by interactions, 
we study the sensitivity of the ground state to a change in the 
boundary conditions \cite{letter1}. 
Boundary conditions are always taken periodic in the transverse 
$y$-direction, and such that the system encloses an 
Aharonov-Bohm flux $\phi$ in the longitudinal $x$-direction. Imposing 
$\phi=\pi/2$ ($\phi=\pi$ corresponds to anti-periodic boundary 
conditions), one drives a persistent current of total longitudinal 
and transverse components given by 
\begin{equation}
\label{current}
I_{l}=- \frac{\partial E_0}{\partial \phi} 
=\frac{\sum_i I_i^l}{L}
\,\,\,\,\mbox{and}\,\,\,\, 
I_{t}=\frac{\sum_i I_i^t}{L},   
\end{equation}
with $E_0$ ground state energy. 
The local current $I_i^l$ flowing at the site $i$ in the longitudinal 
direction is defined by $I_i^l =2 {\rm Im} \langle \Psi_0 | 
c^{\dagger}_{i_{x}+1,i_y} c^{}_{i_x,i_y} | \Psi_0 \rangle$, with an    
analogous expression for $I_i^t$ ($|\Psi_0\rangle$ is the ground state 
wavefunction).  
The response is paramagnetic if $I_{l}>0$ and diamagnetic if $I_{l} < 0$. 
Exact diagonalization techniques (Lanczos method) are possible only 
for small system sizes and here we consider $N=4$ particles in 
$L^2=36$ sites, for $W=5,10,15$ ($k_F l=2.7,0.7,0.3$ respectively) 
and $0\leq r_s \leq 42$. We summarize in Fig. \ref{fig1} results 
from a statistical study of an ensemble of $10^3$ clusters. 
$|I_l|$ and $|I_t|$ have acceptable log-normal distributions for 
all values of $r_s$ when $W\geq 5$ \cite{letter1} and therefore
the log-averages shown in Fig. \ref{fig1} can be considered as typical 
values of the persistent currents:  
$I_{l,\mbox{typ}}=\exp<\ln|I_{l}|>$, with an analogous expression 
for $I_{t,\mbox{typ}}$ (brackets indicate ensemble average).  
Fig. \ref{fig1} (a) shows that the ground state transverse current 
is suppressed before the longitudinal current. 
It is then possible to single out an intermediate phase, in which 
the local currents flow in an ordered way along the direction imposed 
by the flux \cite{letter1,berkovits}. This behaviour is evident in  
Fig. \ref{fig1} (b): the ground state local current angles 
$\theta_i=\arctan(I_i^t/I_i^l)$ undergo a transition from a 
nearly uniform angular distribution (glass of currents randomly 
directed in the plane) towards a distribution strongly peaked in the 
longitudinal direction. As a consequence, the sign of the magnetic 
response becomes independent of the microscopic realization of the 
random potential (see Fig. \ref{fig1} (b) insert). 
Notice that, for a not too strong disorder ($k_F l >1$), in the 
intermediate phase $I_{l,\mbox{typ}}$  
is enhanced by interactions up to a factor $3$ 
(for $W=5$, $r_s\approx 6$) before being suppressed due 
to charge crystallization \cite{letter1}. 
Fig. \ref{fig1} (c) displays the behaviour of the longitudinal  
current for the $20$ lowest energy levels. The enhancement in  
$I_{l,\mbox{typ}}$ disappears away from the ground state. 
In addition, it is possible to 
identify three different regimes when $W=5$: for 
approximately $r_s<2$ and  
$r_s>17$ the longitudinal current increases with the excitation 
energy $\epsilon$, all the contrary for intermediate $r_s$ values. 
We remind that experimentally the metallic (insulating) 
phase has been identified \cite{kravchenko,yoon,hamilton}  
from a decrease (increase) with temperature 
of a related transport property, the conductance.  
For $W=15$, $I_{l,\mbox{typ}}$ always 
decreases with $\epsilon$ (Fig. \ref{fig1} (c) insert), 
indicating that a too strong disorder 
probably forbids a metallic phase in the thermodynamic limit. 
Fig. \ref{fig1} (d) shows that, increasing $\epsilon$, the 
transverse current is suppressed at larger $r_s$ values, eventually 
compatible with $r_s$ values at which the longitudinal current is 
suppressed. This fact suggests that the intermediate phase 
is destroyed at large enough temperatures. For $\epsilon>2$ 
(the Fermi energy is given by $\epsilon_F=\pi n_e t = 0.35$) the sign 
of the persistent current is disorder dependent at any $r_s$ value
(Fig. \ref{fig1} (d) insert). 

\section{Finite size scaling for the $N$ body localization length}  

\begin{figure}
\centerline{\epsfxsize=13cm\epsfysize=13.cm\epsffile{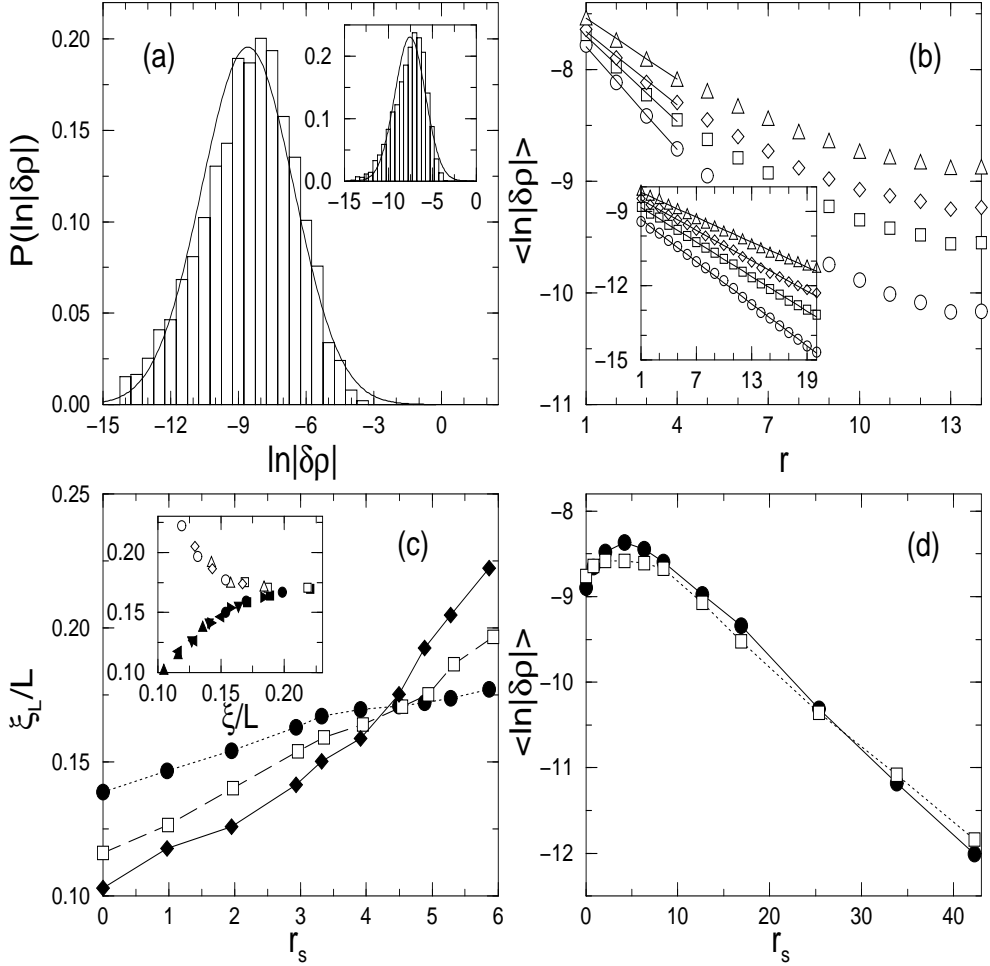}} 
\caption{
(a) Distribution of the logarithms of the charge density variation  
$|\delta\rho|$ for $L=31$, $r_s=0$, $W=10$, $r=5$, fitted by a 
log-normal function of mean $\mu=-8.6$ and variance $\sigma^2=4.2$. 
Insert: the same at $r_s=5.9$, with $\mu=-7.5$, $\sigma^2=3.0$.  
(b) Log-average of $|\delta\rho|$ as a function of $r$, for 
$L=28$, $W=10$, $r_s=0$ (circles), $r_s=2.0$ (squares), 
$r_s=3.9$ (diamonds), and $r_s=5.9$ (triangles). 
The lengths $\xi_L$ are given by the inverse slopes of the straight 
lines. 
Insert: the same for a quasi-one dimensional geometry ($L_x=80$, 
$L_y=8$) at $W=8$ (ensemble of $10^3$ clusters). 
(c) Reduced localization length $\xi_L/L$ as a function of $r_s$ 
for $W=10$, $L=24$ (circles), $L=28$ (squares), $L=31$ (diamonds). 
Insert: ratios $\xi_L/L$ mapped onto the scaling curve $f$ as a 
function of the reduced scaling length $\xi/L$. 
Localized branch (filled symbols): $r_s=0$ (triangles up), 
$r_s=1.0$ (triangles left), $r_s=2.0$ (triangles down), $r_s=2.9$ 
(triangles right), $r_s=3.3$ (circles), $r_s=3.9$ (squares). 
Delocalized branch (open symbols): $r_s=4.5$ (squares), 
$r_s=4.9$ (triangles), $r_s=5.3$ (diamonds), $r_s=5.9$ (circles). 
(d) Log-average of $|\delta\rho|$ as a function of $r_s$ for $N=4$ 
particles in a $6\times 6$ lattice, $r=1$, $W=5$ (circles) and 
$W=15$ (squares) (ensemble of $5\times 10^2$ clusters). 1B states 
are filled in starting from the lowest energy one. 
}   
\label{fig2}
\end{figure}

Though small clusters exhibit an enhancement of $I_l$ for intermediate 
$r_s$, exact diagonalization does not allow to vary system size and 
to establish if the intermediate phase is metallic at the thermodynamic
limit. In order to answer this question and to understand how Coulomb 
interaction destroys Anderson localization, we extend 
\cite{letter2} finite size scaling analysis \cite{pichard,kramer} 
to interacting systems. 
We consider an ensemble of $5\times 10^3$ clusters with $N=3,4,5$ particles 
in square lattices of size $L=24,28,31$ respectively, corresponding to 
very low filling factors $n_e\approx 5\times 10^{-3}$. 
To have Anderson localization inside these     
sizes we consider a large disorder to hopping ratio  
$W/t=10$. Therefore the low energy tail of the one body (1B) spectrum 
is made of impurity states trapped at some site $i$ of exceptionally 
low $v_i$. As we are interested in studying the effect of Coulomb 
repulsion on genuine Anderson localized states we get rid of the 
band tail. Typically we ignore the $L^2/2$ first 1B levels   
(but results do not change provided that the Fermi level if out 
of the band tail, $\epsilon_F>-4t$; under this condition one can 
roughly estimate $k_F l\approx 1$). From this restricted subset of 1B
states we build a basis for the N body (NB) problem, truncated to the 
$N_H=10^3$ Slater determinants of lowest energy (convergence tests 
are discussed in \cite{letter2}).  
The scaling ansatz for the NB problem reads 
\begin{equation} 
\label{ansatz} 
\frac{\xi_L}{L}=f\left(\frac{L}{\xi}\right),  
\end{equation}
where we assume it is possible to map the localization length 
$\xi_L$ at the system size $L$ onto a scaling curve $f(L/\xi)$, 
with $\xi$ characteristic scaling length of the infinite system.   
To characterize the NB ground state by a suitable localization 
length $\xi_L$, we consider the change $\delta \rho_j$ of the charge density 
induced by a small change $\delta v_i$ of the random potential $v_i$ located 
at a distance $|i-j|$. 
To improve the statistical convergence, 
we calculate more precisely the change $\delta \rho (r) = 
\sum_{j_y} \delta \rho_{r,j_y}$ of the charge density on the $L$ 
sites of coordinate $j_x=r$ yielded by the change $v_{0,i_y} 
\rightarrow 1.01 v_{0, i_y}$ for the $L$ random potentials 
of coordinate $i_x=0$. 
Fig. \ref{fig2} (a) shows that $|\delta \rho(r)|$  
is reasonably fitted by a log-normal distribution. Therefore it  
makes sense to characterize the typical strength of the fluctuations by
$\delta \rho_{\mbox{typ}} (r) = \exp <\ln | \delta \rho (r)| >$ 
and extract the length $\xi_L$ over which the perturbation is effective 
from the exponential decay $\delta \rho_{\mbox{typ}} (r) \propto  
\exp(-r/\xi_L)$. 
Such a decay occurs only over a scale $ r << L/2$ since the boundary 
conditions are periodic.  In Fig. \ref{fig2} (b) the lengths $\xi_L$ 
are obtained from the slope of the linear parts of the curves 
(straight lines). A good exponential decay on a much larger interval 
can be observed in a quasi one dimensional geometry ($L_x\gg L_y$).  
Fig. \ref{fig2} (c) gives how the reduced localization length $\xi_L/L$  
depends on $r_s$ for the three considered sizes. A critical point 
appears at $r_s^F\approx 4.3$: for $r_s<r_s^F$,  
$\xi_L/L$ decreases with $L$ (Fermi glass) while it increases for 
$r_s>r_s^F$ (Coulomb metal). 
Fig. \ref{fig2} (c) insert verifies the scaling ansatz 
(\ref{ansatz}): assuming suitable
scaling lengths $\xi$ all the data can be mapped onto a 
single scaling curve $f$, which develops two branches, as typical 
of second order phase transitions (like the 1B Anderson transition in 
three dimensions \cite{pichard,kramer}). 
A power fit of $\xi\propto |r_s-r_s^F|^{-\nu}$ yields a rough estimate 
for the critical exponent $\nu\approx 4$. 
In the quasi one dimensional case we were able to detect a strong 
interaction induced enhancement of the localization length 
(see Fig. \ref{fig2} (b) insert) but no signature of a phase transition.  
Fig. \ref{fig2} (d) shows that a local perturbation in the random potential 
plays a role similar to a change in the boundary conditions. For the small 
system sizes accessible to exact diagonalization it is not possible to 
extract the length $\xi_L$. However, it is possible to see how the 
typical value $\delta\rho_{\mbox{typ}}(r)$, evaluated for example when 
$r=1$, evolves with $r_s$. This quantity increases in the intermediate 
phase before decreasing due to charge crystallization. 
Unfortunately it is not possible, within the approximate technique 
described above, to evaluate how $\xi_L$ changes with $L$ in 
the Wigner crystal phase.  

\section{Conclusions}

Coulomb repulsion can drive a two dimensional system of  
spinless fermions in a random potential towards a new metallic phase, the 
Coulomb metal, which is different from the Fermi glass and the Wigner 
crystal. The local current $I_i$ could be the order parameter 
driving the Fermi glass-Coulomb metal transition (through $\theta_i$) 
and the Coulomb metal-Wigner crystal transition (through $|I_i|$). 

\vspace*{0.25cm} \baselineskip=10pt{\small \noindent This work is 
partially supported by the TMR network ``Phase coherent dynamics of 
hybrid nanostructures'' of the EU.}
%
%
%
%
%
%
%
%
%
%
%
%

\end{document}